# Data-driven recommendations for enhancing real-time natural hazard warnings, communication, and response

Kate R. Saunders[1-3*], Owen Forbes[2], Jess K. Hopf[4], Charlotte R. Patterson[2,5], Sarah A. Vollert[2,6], Kaitlyn Brown[2,6], Raiha Browning[7], Miguel Canizares[2], Richard S. Cottrell[8-10], Lanxi Li[2,11], Catherine J.S. Kim[2,13], Tace P. Stewart[2,6], Connie Susilawati[2,12], Xiang Y. Zhao[5], Kate J. Helmstedt[2,6]

**Affiliations:**
[1] Department of Econometrics and Business Statistics, Monash University, Victoria, Australia
[2] Centre for Data Science, Queensland University of Technology, Brisbane, Australia
[3] ARC Centre of Excellence in Climate Extremes, Monash University, Australia
[4] Hatfield Marine Science Center, Oregon State University, Newport, Oregon, USA
[5] School of Biology and Environmental Science, Queensland University of Technology, Brisbane, Australia
[6] School of Mathematical Sciences, Queensland University of Technology, Brisbane, Australia
[7] Department of Statistics, University of Warwick, Coventry, UK
[8] Institute for Marine and Antarctic Studies, University of Tasmania, Hobart, Australia
[9] Centre for Marine Socioecology, University of Tasmania, Hobart, Australia
[10] School of the Environment, The University of Queensland, Brisbane, Australia
[11] School of Computer Science, Queensland University of Technology, Brisbane, Australia
[12] School of Economics and Finance, Queensland University of Technology, Brisbane, Australia
[13] School of Earth and Atmospheric Sciences, Queensland University of Technology, Brisbane, Australia

**Acknowledgements:**
The ideas of this paper were first identified during a hackathon run in rapid response to the flooding that affected Brisbane, Australia in February of 2022. This hackathon aimed to capture the ephemeral nature of data during a natural disaster event and characterise and review the real-time response. This paper is therefore motivated by the authors' own experiences during the flooding and provides the unique dual perspective of both experienced data practitioner and firsthand witness. Thank you to the Centre for Data Science at the Queensland University of Technology for their support running the hackathon and to all the participants.

## Abstract

The effectiveness and adequacy of natural hazard warnings hinges on the availability of data and its transformation into actionable knowledge for the public. Real-time warning communication and emergency response therefore need to be evaluated from a data science perspective. However, there are currently gaps between established data science best practices and their application in supporting natural hazard warnings. This Perspective reviews existing data-driven approaches that underpin real-time warning communication and emergency response, highlighting limitations in hazard and impact forecasts. Four main themes for enhancing warnings are emphasised: (i) applying best-practice principles in visualising hazard forecasts, (ii) data opportunities for more effective impact forecasts, (iii) utilising data for more localised forecasts, and (iv) improving data-driven decision-making using uncertainty. Motivating examples are provided from the extensive flooding experienced in Australia in 2022. This Perspective shows the capacity for improving the efficacy of natural hazard warnings using data science, and the collaborative potential between the data science and natural hazards communities.

## Keywords
Natural hazards, natural disasters, data science, data visualisation, warnings

## Introduction

Climate change is increasing the frequency and intensity of many extreme weather events[1]. Combined with changes in land use and higher urbanisation rates, this means that communities are being increasingly exposed to natural disasters like floods and fires[2]. Natural disasters cause severe and negative impacts to communities, encompassing socio-economic[3], health[4] and environmental impacts[5]. The insurance costs can be billions of dollars[6,7]. However, people continue to live in at-risk areas[8–10]. Many communities in these areas are uninsured, under-insured or uninsurable against future inevitable disaster events[11]. Since communities are at risk from natural disasters, there is a clear role for policy and planning in improving disaster preparedness, resilience, and response.

One pathway to reducing community impacts is through better use of available data for forecasts and warnings. In the current digital age, natural hazards related data is more accessible than ever before. Communities have never been more connected virtually, and the world has never been more observed. Globally, satellites capture images of the land from space at increasingly fine spatial and temporal scales[9,12]. Locally, citizen scientists collect and upload data in real-time[13], and other members of the public record observations of their surroundings and upload them to social media[14,15].

High data availability has facilitated a massive growth in observations and our understanding of natural and human systems. This creates opportunities for data science to address modern challenges in natural hazards science, communication, and policy. A growing body of research is therefore demonstrating the potential for data science to aid in disaster management, with emphasis on areas such as big data[16,17], novel data sources[14,18], and information systems[19,20]. However, within natural hazards operational practice, data science is still under-utilised. Particularly, for supporting real-time warnings, communication, and emergency response (Figure 1).

Our objective is to identify major opportunities for data science to enhance real-time natural hazard warnings. We delve into the complexities of modern hazard warning systems and explore how data science can help address existing challenges. This involves examining how data are used at key stages in the information pathway that connects (i) observations, (ii) weather forecasts, (iii) hazard forecasts, (iv) impact forecasts, (v) warnings and (vi) decision making processes, known as the warning-value chain[21,22].

This perspective paper is organized as follows. It begins with a discussion of visualisation practice and needs for hazard forecasts. Followed by an explanation of barriers to effectively using data for impact forecasting. From barriers we look to opportunities, exploring how data could be used to support highly localised and more user-centric forecasts. Finally, we stress the importance of uncertainty in forecasting and warning communication. We use examples from the widespread flooding that occurred across multiple states in Australia during 2022 to highlight limitations in current practice and possible opportunities for innovation (Fig. 1). We identify 4 priority actions to encourage collaboration between the communities of data science and natural hazards science. By illustrating opportunities for future collaboration, we outline a pathway toward improved data-driven hazards communication.

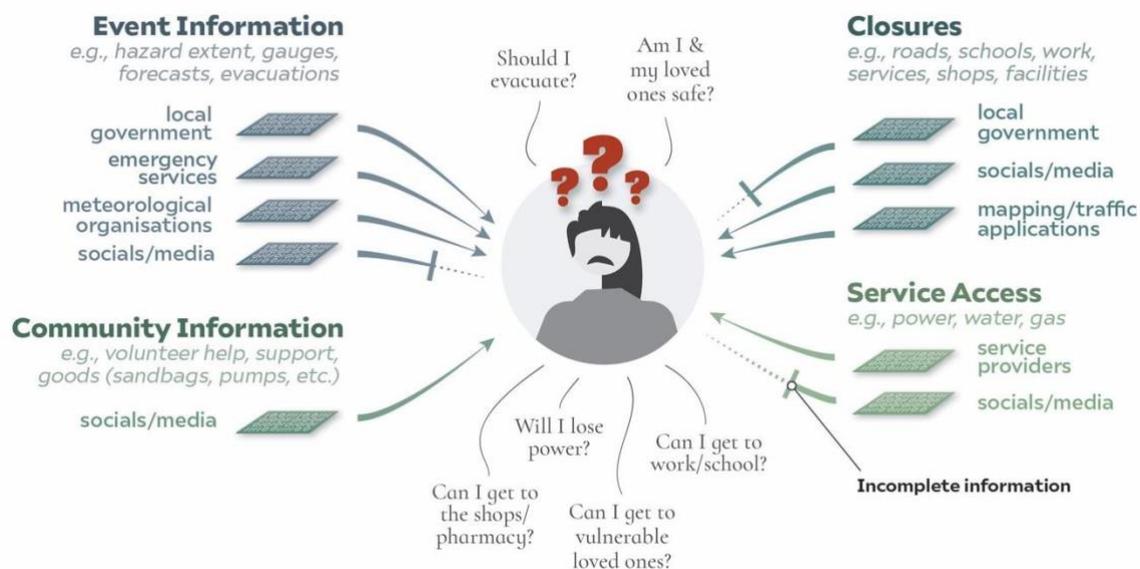

**Figure 1:** An idealised example showing a user-centric perspective on the vast, varied and at times incomplete data available for decision making during a natural hazard. Individuals are confused about how best to convert this data into information about their exposure, impacts and vulnerability to make decisions under stress. For firsthand community accounts of the Eastern Australian in 2022 flooding refer to Natural Hazards Research Australia report[23].

# Best-Practice Data Visualisation for Hazard Forecasts

A hazard forecast provides details of the expected danger to the public resulting from a physical weather event. These forecasts typically communicate the extent of a hazard event or the degree to which humans, property, and the environment will be affected by the event as it unfolds [24,25]. Hazard forecasts are informed by observational data and weather forecasts[21]. For example, a weather forecast predicting extreme rainfall and river gauge observations would be used to produce a hazard forecast showing flood prone areas.

Hazard forecasts are vitally important for facilitating an emergency response. However, the effectiveness of such forecasts for initiating appropriate actions by the public is highly dependent on how we choose to visualise forecasts. Visual communication of hazard forecasts during an emergency can fail to adhere to best-practice data visualisation principles[26]. We outline and expand on these principles below, to ensure that information visualised in hazard forecasts are relevant, interpretable, and actionable.

**Adhere to Data Visualisation Standards**

Hazard visualisation falls within general scientific visual communication - which has a long, established history to draw upon[27]. Hazard visualisations must therefore emphasise essential information through a careful use of colour, contrast, and text[28,29] and must effectively convey the magnitude of a hazard to trigger an appropriate response[30]. This includes using recognisable, culturally relevant colour schemes in emergency scenarios to promote an intuitive understanding of the variables displayed, such as yellow-to-red for escalating hazards, and sequential blue colour scales for flooding depths[29,31]. Colour scales should always be accompanied by a clear and simple legend[29]. Visualisations should consider accessibility for people with low vision or colour blindness, for example by employing accessible colour palettes, or using contrast to delineate layers[31]. The scope of the hazard forecast should be clearly labelled in the visualisations as the forecasts are widely shared and may be separated from their source. This scope includes the time period for which the hazard map is relevant, the prediction sources, and whether the visualisation is modelled or observed[31]. An important difference between hazard visualisation and general scientific visualisation is that people's capacity to process information may be reduced in an emergency, so this needs to be a key consideration in any hazard visualisation design[32,33].

**Allow People to Tailor and Control Their Warning Information Interactively**

Warnings are more actionable when viewers can place hazard forecasts within their personal context. Interactive maps are a key method to support meaningful interpretation of hazard forecasts. They leverage the public's growing digital literacy by allowing viewers to explore predictions through features they are accustomed to, such as zooming, panning, and searching for specific addresses[34]. Interactive elements can also help reduce the cognitive load induced when viewing complex data on a single screen[35,36] and increase understanding of personal risk[37]. Overlaying flood extent predictions on satellite imagery and including 3D viewing capabilities (e.g., Google 'Street View') can further improve user orientation and impact assessment[31,34,38]. Despite clear public demand for interactive mapping and growing digital literacy[39], issuing static hazard maps is still commonplace in real-time hazard communication.

**Allow Flexibility and Cater to Different Target Audiences**

Hazard visualisations need to be designed to effectively communicate with multiple different audiences. Visualisation design needs to be simple to support emergency decision-making, while also balancing the broader public desire for more-detailed and user-centric hazard information[28]. Given the wide variation in different users' needs, contexts, and information literacy, this presents challenges to presenting a single simple, accessible visualisation. Multiple different visualisations may therefore be needed to help the public to understand different aspects of their hazard exposure and the decision-making process it necessitates[40,41]. This is where interactive maps can be used to great effect. Layering the simplest and most pressing information first, followed by optional levels of additional detail. Interactive maps can therefore present multiple lines of evidence within a single format, flexibly supporting different audiences' needs.

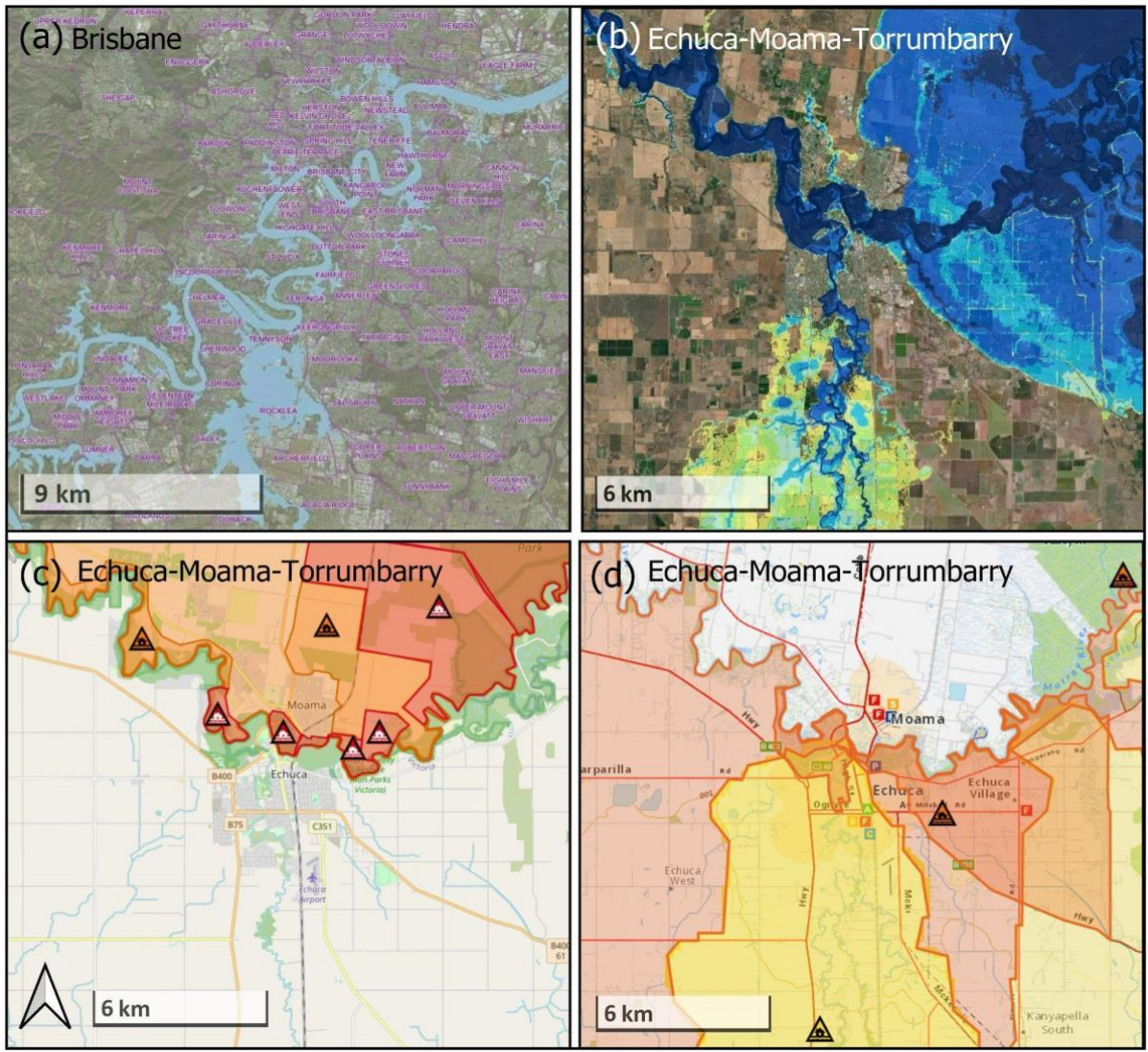

**Figure 2:**[1] Screenshots of hazard forecasts shared with the Australian public during different flood events: (a) February 2022 flooding in Brisbane, Queensland (QLD) and (b-d) October 2022 flooding on the state border of New South Wales (NSW) and Victoria (VIC) impacting the extended Echuca-Moama-Torrumbarry region. These figures show: (a) A static pdf map of predicted maximum flood extent for Brisbane[42]; (b) Online interactive map showing the Echuca-Moama-Torrumbarry predicted flooding extent without mitigation[43] (c) – (d) Online interactive maps showing minor, moderate, or major flood warnings using polygons, (c) shows NSW warnings[44] and (d) Victorian warnings[45]. Polygons maps in (c) and (d) increase in resolution with zoom and have pop-up information that appears when clicking the screen. Map legends were present on all sub-figures but are cropped from view. Effective colour scales were present for (b-d).

---

[1] Figures (a) and (c) are covered under a Creative Commons Licence 4.0. Figures (b) and (d) are available under Australian Copyright law and Fair Dealing for research, which has provisions for study, criticism, and review.

**Box 1: Visualisation of Hazard Forecasts**

Here we unpack several examples of hazard forecasts provided during the Australian floods of 2022, shown in Figure 2. We consider the extent to which these visualisations adhere to best-practice principles and support emergency decision-making.

Maps in the top row of Figure 2 show the predicted maximum flood extent for floods in (a) the Brisbane region, Queensland and (b) the Echuca-Moama-Torrumbarry region, on the border of New South Wales and Victoria.

Figure 2 (a) was the only hazard visualisation provided by local government to the public during the Brisbane flooding to support real-time decision making. However, map (a) did not implement the best-practice principles discussed above. It is a static map, without the ability to zoom or obtain localised information about hazard exposure. No depth information was provided. Map (b) in contrast did offer these features. Neither (a) nor (b) offer information on event timing, displaying only peak predicted extent. Information about the time of first hazard exposure and expected duration of hazard exposure can assist the public in their decision-making.

Maps (c) and (d) in the second row use interactive polygons coloured by warning levels, enabling clear, simple communication about the predicted hazard. These maps come from interactive visualisation platforms that include a search bar for addresses and additional warning information, such as the expected number of building stories to be inundated. Flashing polygons signal immediate exposure risks and help trigger necessary evacuations upon arrival at the website. Flashing icons effectively draw attention to crucial information. This demonstrates the substantial scope for use of interactive mapping features to engage user attention and facilitate decision making[46,47]. In Queensland, a post-event recommendation has been made to use warning polygons as a minimum standard of hazard visualisation[48].

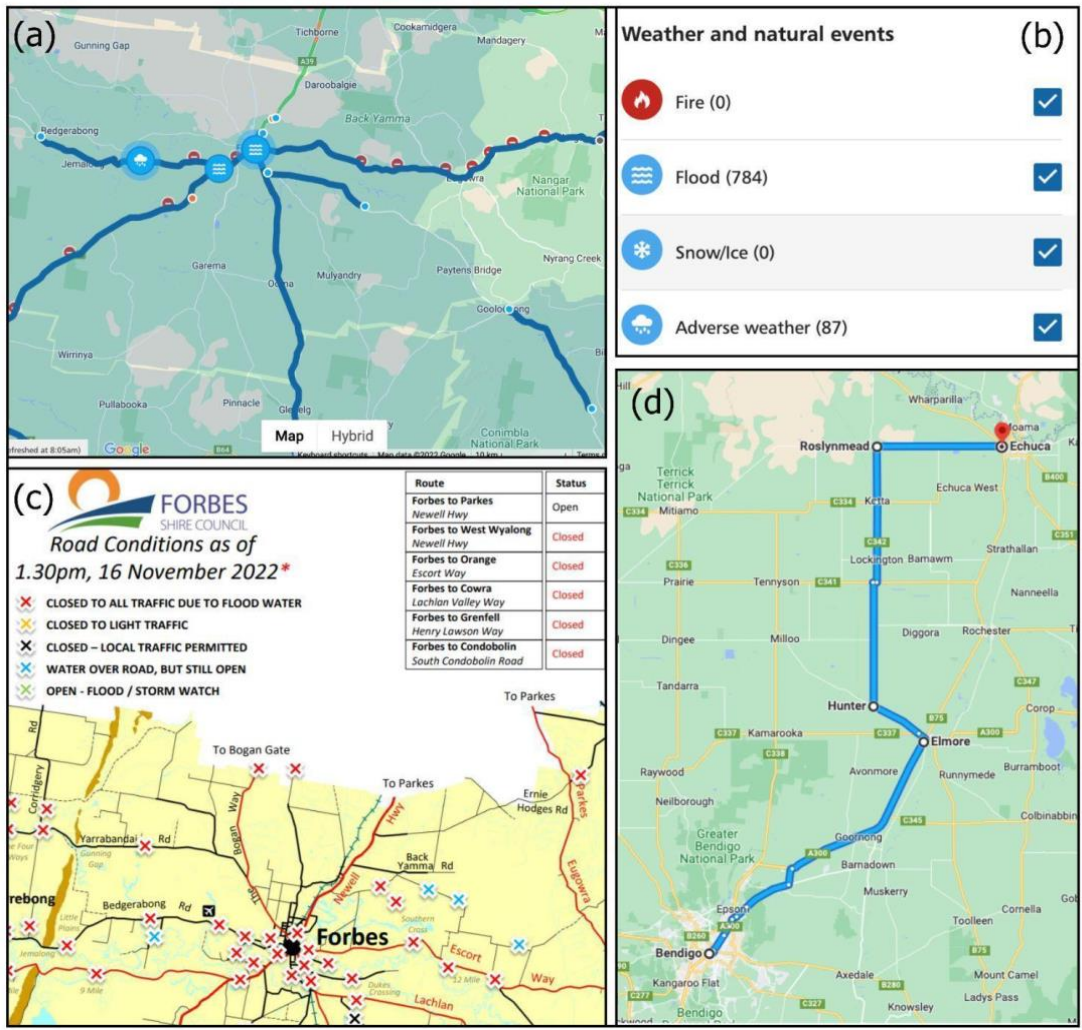

**Figure 3:**[2] Visualisations of impacted roads during different flooding events in Australia in 2022. (a) Screenshot from Live Traffic NSW for the city of Forbes, NSW on 16th of November 2022 when the city was cut into three by flooding. Blue markings show road closures[48]. (b) Screenshot highlighting the number of road closures for different hazards for an extended region of figure (a) on the Live Traffic NSW app[49]. (c) A static pdf map showing local closures in Forbes, NSW[50]. People were referred to this map from Live Traffic NSW, as the online map in (a) was known to not accurately reflect current closures. This shows the importance of local, current on-the-ground knowledge during an event to facilitate evacuations. (d) A screenshot of alternative road closure visualisation from the Echuca, Victoria flooding in October of 2022. Here a Google map link was provided in the warning information showing an open route to the nearest evacuation centre. Map (d) highlights open and safe roads opposed to closures[45].

---

[2] Figures (a) and (b) are covered under a Creative Commons Licence 4.0. Figures (c) and (d) are available under Australian Copyright law and Fair Dealing for research, which has provisions for study, criticism, and review.

**Box 2: Communication and Visualisation of Impact Forecasts**

Here we unpack several examples of impact-based communications and visualisations provided during the Australian floods of 2022. We highlight gaps in the existing approaches to communicating these impacts.

To support communities to understand their impacts from a hazard, there are many web applications ('apps') available. This includes specific hazard-based apps for warning communication [e.g., 44,45] and various disaster dashboards [e.g., 51–53]. However, many of the apps people use during an emergency are not bespoke to natural hazards, but are instead designed primarily for day-to-day use (e.g., Google Maps). During an emergency, the role and importance of apps to support decision making can be very different to everyday scenarios. Natural hazards commonly disrupt normal system operation. Therefore, for apps to operate robustly and support decision making during emergency situations, app functionality and design need to be specific to warning communication of the hazard and impact forecasts.

Consider apps showing road closures. There are three key differences between normal use of these apps and the required functionality during in an emergency. First, under normal use, these apps share information about a handful of closures due to road works. In a natural hazard event, closed roads can outnumber open roads. This changes the intended design and balance of the visualisation. Second, road closure apps work well when it is known whether the road is closed or open. During a natural disaster, a road can change rapidly from safe and open to impassable, which can significantly alter evacuation routes. Data on closures may also be unknown or out of date. Road closure apps are not designed to communicate uncertainty or forecasted changes in the road network with a changing hazard. Third, the app function is not designed to show isolated regions due to impassable roads.

These design gaps can pose life threatening impacts for people who need to make decisions about evacuating or isolating in place. These criticisms can be made of all state-based traffic applications in Australia and yet these apps were still referenced in official warning communication during hazard events (Fig 3). Figure 3 shows different visualisations used to communicate road closures to the Australian public during flood hazards in October and November of 2022.

# Data Challenges for Effective Impact Forecasts

Hazard and impact forecasts are different. Impact forecasts build upon hazard forecasts by including social, environmental, essential service and infrastructure information into the warnings to help communities understand how they could be affected[25,54]. For example, an impact forecast builds upon a hazard forecast for predicted flood extent by providing information about which roads will be cut-off and whether a person might lose power. The translation from a hazard into impact requires a variety of data from multiple sources. In this section, we discuss the associated data challenges to issuing an impact forecast and areas for improvement.

**Diversity in Impacts**

Hazards can result in direct life-threatening impacts, and through changed access to key infrastructure and essential services, indirect life-threatening impacts. There is therefore a huge diversity among the impacts that people can experience stemming from the same hazard (e.g., Table 1). An impact from a hazard can also affect people differently depending on their individual vulnerability and resilience[55–57]. Therefore, when hazards are visualised and communicated, impacts need to be visualised and communicated too (e.g., Fig. 3).

**Table 1:** This table provides key infrastructure and essential services that can be impacted and whose access may change due to being exposed to a hazard (a non-exhaustive list). Examples are provided from the 2022 Southeast Queensland floods that affected Brisbane[48,58]. During these floods, 13 lives were lost and over 500,000 people were affected by flood related impacts. An estimated 30,000 homes, businesses and vehicles were damaged, and 20,000 householders also placed claims resulting from the loss of an essential service[48].

| Type | Potential impacts on individuals and communities | 2022 Brisbane Flood examples |
| --- | --- | --- |
| *Service utility outages* eg. Water, sewerage, power, phone, internet. | Concerns about overflowing sewerage and access to safe drinking water. Changed access to community and family support. Reduced access to digital information and announcements with loss of power. Loss of perishable foods and inability to store perishable foods. | 180,000 customers lost power for multiple days. |
| *Transport outages* eg. Road closures, public transport outages, disruptions and outages of rideshare services. | Isolation of communities (an issue for those already in areas with low access to services). Changed access to essential services such as shops, pharmacies, and healthcare. | 1,400 km of roads closed or restricted, including multi-day closures of the major roads such as the Bruce and Warrego Highways and Ipswich Motorway. 19 Brisbane ferry terminals were damaged, with 6 requiring major |

|  | Changed access to work (an issue for those in critical or service roles) and for loss of income. Changed access to school (an issue for parents still needing to work). Uncertainties about safest travel options during and immediately after the event. | repairs. Ferry terminals were closed and one ferry sank. 60% of bikeways in the Brisbane CBD were inundated. |
|---|---|---|
| **Emergency services** eg. Changed access to police, fire, ambulance. | Longer wait times for emergency and related non-emergency services. |  |
| **Other essential services** eg. Changed access to healthcare, hospitals, pharmacies, and supermarkets, including supply disruptions. | Vulnerable individuals isolated from regular care and support services. Poorer health outcomes for those who cannot access required medical care. Supply disruptions, especially for fresh produce. | 17,000 individuals at the Community Recovery Hubs looking for support. 2,500 primary producers lost 30% of production. |
| **Support services** eg. Changed access to supports for mental health, domestic violence, and drug and alcohol abuse. | Trauma from the hazard event may trigger mental health or prevent access to drug and alcohol supports. Inability to escape from domestic violence (can worsen due to hazard). | The Department of Communities, Housing and Digital Economy undertook 22,000 psychological first aid visits. |

**Interconnected Impacts and Dataset Interoperability**

Impacts from a hazard are inextricably linked. For example, an individual's ability to understand impacted roads is vitally important for understanding how their access to other essential services and key infrastructure has changed, and how these access changes might impact their vulnerability. Datasets therefore need to be combined and analysed jointly for data-driven tools and apps to deliver timely and effective impact forecasts (Fig. 4).

Currently for the public to piece together the data story and impact-based information relevant to them, they are jumping between multiple different websites, news sources, social media platforms and other information providers (Fig 1). This ad hoc exploration is being undertaken despite the difficulty of decision making during natural disasters[32,33] and unequal data literacy across community members[59]. Those who are most vulnerable are also the least likely to be able to acquire and synthesize information about how the hazard may impact them.

Impacts typically require multiple datasets to analyse, such as hazard, spatial, demographic, and essential services data. There is therefore a clear need for data interoperability in order to properly model, visualise and communicate interconnected impacts. Improving data interoperability is also essential to ensuring linkages along the warning-value chain are not broken and that this data pathway is not a leaky pipeline[21,22].

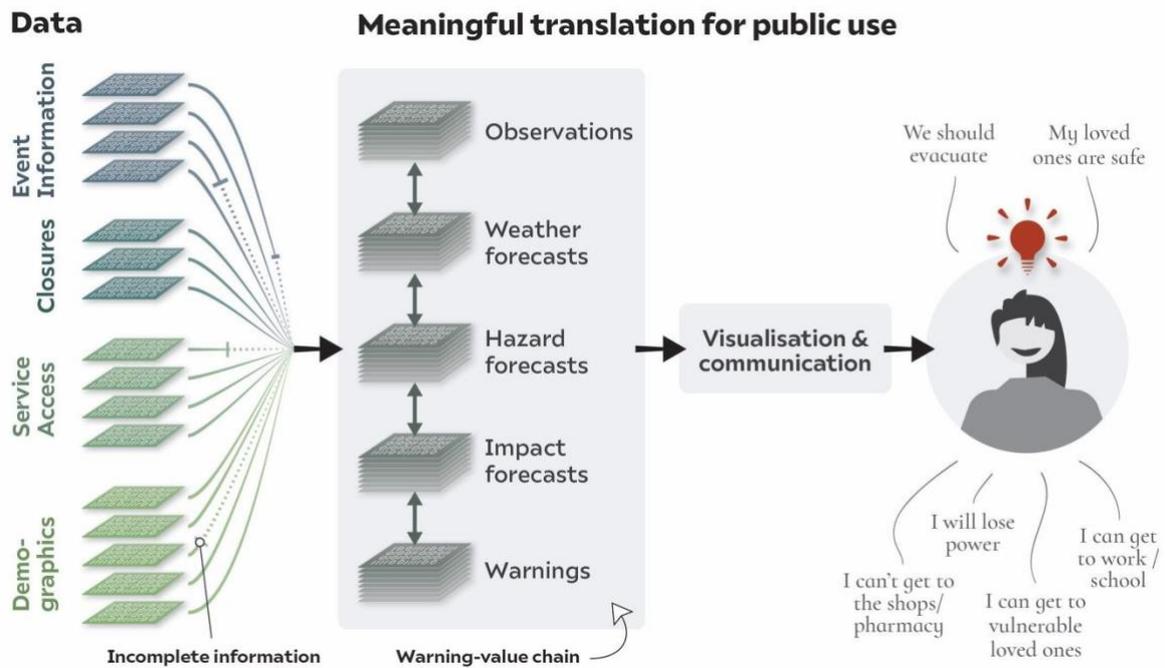

Figure 4: A vision for how improving data interoperability can support more effective warning communication. Swathes of data are needed for hazard and impact forecasting (left panel). These data are needed in various combinations to inform different stages along the warning value chain (middle panel). If these data are interoperable and ready to use for emergency purposes, this will increase the efficacy of the warnings. The warnings are then synthesised for the public and clearly communicated using state-of-the-art data-visualisation. As a result, the public can better understand their exposure and impact, and make more informed decisions (right panel). This contrasts with the data confusion shown in Figure 1.

**Towards Interoperability and Ending Data Silos**

One limiting factor to improved interoperability are data silos. In many instances, the data needed to translate a hazard forecast into an impact forecast does exist. These data may exist at a highly localised property or street scale for particularly well-resourced, metropolitan locations. However, the data may be siloed, held in government or institutional repositories and be inaccessible for public use. Data may also often be held in proprietary or behind paywalls. These silos exist despite interoperability of different datasets being critical for a well-informed natural hazard response[60] (Fig. 4).

Failure to seamlessly share data across government or institutional boundaries hinders emergency response, and ultimately the public's ability to understand their impacts. The geographic impacts of natural disasters do not adhere to government boundaries or institutional remits. For example, in the October 2022 Echuca-Moama-Torrumbarry flooding both sides of a state border were impacted yet the primary hazard and impact warnings were communicated and visualised using different state government-based apps (Fig. 2c, d).

**Advocating for Open, FAIR Data and for Model Transparency**

Open data sharing benefits emergency management and policy makers who are coordinating a disaster response across multiple different areas and organisations. Open data also benefits data scientists who are looking for ways to improve the real-time emergency response and propose innovative solutions to existing gaps.

Operating transparently around the data and models means that we can better share information during an event, and better evaluate how timely and effective an emergency response was after the event. Such data and model transparency can promote accountability and reduce the chance of misplaced distrust from end users (e.g., the public)[61–63]. Transparency also supports decision makers when faced with incomplete or inconsistent data[60], and creates opportunities to learn from others' mistakes and successes. It is especially important to learn from those who have already responded to similar natural hazards, both domestically and internationally.

We need to move towards storing data according to the highest open data standards. We should work towards a 5 star open data rating; which means ensuring our data (1) has an appropriate open data licence, (2) are stored in machine-readable, structured formats (3) are stored in non-proprietary formats, (4) has an identifier so the data can be referenced and (5) include links to other reference datasets for ease of automatic integration[64]. Extending these ideas, the data should also be FAIR; easily Findable, Accessible, Interoperable, and Reusable[60,65]. Finally, particularly for emergency purposes, data need to be accompanied by appropriate metadata, including the dates of the last update or dates of any data maintenance. Adhering to these best-practice open data standards greatly enhances the usability of existing datasets available for hazard and impact forecasting.

**Preventing Disappearing Data and Prioritising Reproducibility**

The current lack of reproducibility means our ability to assess the effectiveness of any warning communication is limited. Data disappears at every stage of the warning value chain. Data that are available during the event often disappears without being publicly archived; this includes hazard and impact forecasts (Fig. 2 and 3), observations of on-the-ground conditions, warnings, and data on public adherence to warnings. The risk of data disappearing also increases with time since the disaster event. This represents a significant loss of information that is crucial for understanding how a natural hazard evolved and how effectively it was responded to.

It should be a priority that any data and models used to generate the forecasts are preserved permanently in a fully reproducible manner. This should include information on the warning message timing and content, and any data visualisations associated with the warning communication. This will enable evaluations of whether the warnings and communication adequately supported the decision-making of emergency services and the public during the disaster.

# Data Opportunities for Highly Localised Forecasts

The public increasingly expect individually relevant and targeted warning information. Tailored local data can help people to assume individual responsibility or rely on local, community support during a natural disaster[23]. However, many traditional data sources are not of a suitable spatial or temporal resolution for providing highly localised forecasts. Large spatial gaps in the in-situ observing network mean there may be no localised, on-ground coverage of the extreme event. Many of the available traditional data sources also do not update at the same temporal frequency as that of an evolving hazard situation[66,67]. As a result, state-of-the-art hazard forecasts use polygons to represent affected areas instead of providing more localised information (Fig 2c,d); and the impact information can be out of date with local, on-the-ground conditions (Fig 3a).

Data on a much finer spatial-temporal scale is required for higher resolution forecasts. Technologies such as satellite remote sensing[9,12], offer opportunities to observe the world on a more comprehensive spatial scale[68]. Camera devices and drones[69], also provide a means to collect of data safely and accurately during a hazard event. These new observation technologies can be supplemented with citizen science, including crowdsourcing through websites and apps[13], or social media data[14,15]. Crowdsourced data includes text, image, video, and georeferenced data from mobile devices at critically impacted locations. These technologies can help increase the resolution of the data and consequently the skill of forecasts, such that individuals not only understand the impacts of a flood in their city, but also at the level of their suburb or street.

Crowd-sourced data must be quality assured for real-time emergency purposes due to the importance of accurate information in life-threatening situations. Pre-processing is necessary to handle large amounts of crowd-sourced data and ensure its accuracy[70]. Data wrangling and pre-processing is aided by advances in data science, such as text-mining, web-scraping, machine learning and deep learning[71–73]. In some instances where the crowd-sourced data have a variable or unknown quality, value can still be generated from the data due to the sheer volume creating a distinctive pattern amid the background noise[74].

Meeting the public's expectation for more detailed and highly localised hazard and impact forecasts can only be made possible through increasing the spatio-temporal resolution of available data. Novel, rapidly updating data sources produce large amounts of data at highly localised scale, which are available frequently and quickly. Data at such localised scales provide additional benefits for forecast validation and confirming on-the-ground conditions.

These more novel data sources are a low burden to emergency services but are highly valuable for supporting a real-time response and assisting decision-making under uncertainty. Warning communication and emergency response can therefore be enhanced through increased uptake of these novel data sources, which requires establishing pipelines for (i) processing, (ii) quality assurance and (iii) integration with current methodologies.

# Embracing Uncertainty

**Uncertainty Along the Warning Value Chain**

Weather, hazards, and impacts forecasts always have some underlying uncertainty, but emergency managers must still communicate rapid and urgent recommendations to the public. How uncertainty is propagated along the warning value chain therefore needs to be carefully considered and factored into emergency planning and warning communication. This includes uncertainty in both space and time as the hazard event unfolds.

Uncertainty arises at multiple stages of measurement, data collection, modelling, and analysis. Ensemble forecasts from numerical weather models show multiple possible realisations of how the weather could unfold[75]. Weather forecasts are then used to produce hazard forecasts. If only a summary or a high-resolution run from the ensemble weather forecast is used in the hazard model, this will not account for the full range of possible outcomes[76]. Similarly, if a single hazard forecast is used to generate an impact forecast, then the full range of possible impacts will not be accounted for.

Underestimating the possible risks hinders the ability of emergency management to effectively respond and leaves people vulnerable[77]. Uncertainty therefore should be treated as a rich information source, so one can better assess and comprehend the full range of potential impacts and their interactions. This empowers more effective and timely decision-making. Embracing uncertainty can provide important insights for disaster management, and boost end-users' trust in the accuracy of the information given in a hazard forecasting scenario[78]. Research has shown communicating weather forecast uncertainties and probabilities can help people to make better decisions, increases trust in the forecast, and even improves compliance with warnings[79,80].

Messaging around uncertainty for hazard and impact forecasts must be carefully managed given the inherent difficulty of making decisions and the impaired cognitive capacity under stress, especially in natural disasters[32,33]. However, this does not preclude emergency management from visualising spatio-temporal uncertainty to inform their in-house decision making. Further, clear and easily understandable uncertainty visualisation could help balance the false alarm risk when communicating hazard and impact forecasts with the public, giving people more capacity to make informed decisions.

It is unclear to what extent uncertainty was propagated along the warning chain during the Australian 2022 floods, and how this factored into emergency management decision making and the timing of warning communication with the public. In some instances, directly communicating hazard uncertainty was avoided by using broad warning categories (eg. Fig 2c, d). There was no clear visual communication of uncertainty in the impact forecasts, such as road closures (Fig 3a). In many instances people were given very little warning[23], so potential benefits are large if uncertainty can be better used to improve the timeliness of warning communication. Statisticians can work alongside emergency management to better account for uncertainty in the data and aid decision-making[81].

**Visualising Event Uncertainty**

Communicating a useful level of uncertainty to the public is a challenge for natural hazards. As with any data visualisation, visualising uncertainty needs to be tailored to the intended audience's context, needs and data literacy. For hazard forecasts, there are several variables for which providing uncertainty information could give the public more insight into the forecasts and more understanding in the predictions. These include uncertainty in predicted spatial extent, likelihood of hazard exposure at a given location, possible severity of a hazard at a given location, and the timing window of hazard arrival. This is similar for impact forecasts, where the uncertainty around predicted impacts and their timing and severity needs to be communicated clearly for people to make timely and informed decisions about their individual exposure and vulnerability.

When visualising spatial data such as hazard and impact maps, multiple variables in the data can be visualised simultaneously using attributes such as colour, shading, transparency, spacing, shape or textures[82,83]. Uncertainty can be communicated through these same visual attributes. For example, colour can be used to show predicted hazard exposure at a given location, while transparency could be used to represent the confidence of emergency services that the same location will be directly impacted. This would prevent a false sense of confidence in single point estimates. For instance, the failure to discuss and visualise uncertainty in estimated flood levels, including the 1 in 100 year level or peak maximum flood level, could mislead people into thinking they are safe when in fact their location is within uncertainty for that estimated value.

The level of complexity and detail in uncertainty communication and data visualisation needs to be tailored to the audience. One simple approach to communicate uncertainty includes categorising an uncertainty measure into discrete bins with qualitative labels such as "high" "medium" and "low" uncertainty, with explanatory legends[78]. More sophisticated approaches involve the use of choropleth maps[84]. A blending of different visualisation approaches for uncertainty can be used to control the complexity of the messaging in a visualisation, including blending choropleth maps, binned uncertainty categories and using interactive features[47,84].

# Outlook

Data and data science tools can be better used to support real-time warnings, communication, and emergency response during a natural hazard. There already exist data-driven solutions to many current challenges in warning communication. Integrating data science methods into natural hazards and emergency management frameworks is therefore a priority: There are clear opportunities in this space for the betterment of warning communication.

This perspective paper has assessed data limitations in existing warning communication and provided data-driven recommendations. We highlight four priority actions:

1. Adhere to best practices in visualisation and appropriately integrate interactive elements,

2. Work towards open, FAIR data and transparent models, particularly for hazard and impact forecasts (FAIR: Findable, Accessible, Interoperable and Reproducible[60]),

3. Create infrastructure so that novel data sources, like citizen science data, can be used to support more localised warnings, and

4. Better incorporate uncertainty into emergency decision-making, visualisation, and warning communication.

These priority actions form a starting point for interdisciplinary conversations about how we can better use data and data science to support real-time emergency response and warning communication. This is a conversation that will need to be ongoing and collaborative to ensure that these priority actions are translated into an operational setting.

Adherence to these priority actions should also form an important part of any post-disaster review into the effectiveness and adequacy of warnings. We need to critically evaluate how data was being used along the warning-value chain, and where important data for warnings might have been incomplete, inadequate, missing, or under-utilised (Fig. 4). Any post-event evaluation also needs to consider the effectiveness of data visualisation for supporting warning communication and decision-making.

These priority actions outline the important role data science can play in enhancing real-time natural hazard warnings. In the face of limited funding and resources for natural hazard preparation and response, data science offers cost-effective opportunities. Progress on these priority actions will therefore serve to centre data and data science tools as important solutions critical to future disaster preparedness.